# Effect of ancilla's structure on quantum error correction using the 7-qubit Calderbank-Shor-Steane code

P. J. Salas[(1)], A.L. Sanz[(2)]

[(1)]Dpto. Tecnologías Especiales Aplicadas a la Telecomunicación,

[(2)]Dpto. Física Aplicada a las Tecnologías de la Información,

E.T.S.I. Telecomunicación, U.P.M., Ciudad Universitaria s/n, 28040 Madrid (Spain)

**Abstract**

In this work we discuss the ability of different types of ancillas to control the decoherence of a qubit interacting with an environment. The error is introduced into the numerical simulation via a depolarizing isotropic channel. The range of values considered are $10^{-4} \leq \varepsilon \leq 10^{-2}$ for memory errors and $3.\,10^{-5} \leq \gamma/7 \leq 10^{-2}$ for gate errors. After the correction we calculate the fidelity as a quality criterion for the qubit recovered.

We observe that a recovery method with a three-qubit ancilla provides reasonable good results bearing in mind its economy. If we want to go further, we have to use fault-tolerant ancillas with a high degree of parallelism, even if this condition implies introducing new ancilla verification qubits.

## I. INTRODUCTION.

The field of quantum information has undergone an incessant growing starting at the very moment of recognizing that the quantum treatment of information could provide some advantageous features totally out of reach of the classical treatment. The application of quantum information theory to computers has given rise to the concept of a quantum computer [1] capable of carrying out non-classical algorithms. This ability is based on two specific quantum attributes: parallelism (i.e. the possibility of performing an exponential number of operations with polynomial resources) and interference (i.e. the possibility of controlling the amplitudes of certain terms appearing in the development of a quantum state on an adequate basis). The main advantage of quantum computers compared to classic ones, rests on an adequate hardware allowing the coherence of the quantum states used by the computation to be maintained for a period of time long enough to conclude the algorithm. But it is known that, unhappily, there are many sources of error opposing this intention [2]. Some faults may appear owing to an inaccurate application of quantum gates; other errors come from the interaction of the computer with the environment due to its lack of isolation; there will also be reading errors, and so on. All of them generate a loss of coherence in quantum states (decoherence). Quantum error correcting codes (QECC) have been introduced [3] with the intention of detecting and correcting errors, increasing therefore the coherence time. A general type of QECC is the stabilizer codes [4] in which the vector space of the quantum code is specified by the eigenstates of a set of operators constituting an abelian subgroup from the Pauli group of order n, $G_n$. The operators of $G_n$ are the n-fold tensor products $(\sigma_{i_1} \otimes \sigma_{i_2} \otimes \ldots \otimes \sigma_{i_n})$ of single qubit Pauli operators $\sigma_{i_k}$ where $i_k$ = I, x, y, z.

Error correcting codes are themselves quantum computations prone to inaccuracies, and may introduce new errors. A strategy that works efficiently even when its own errors are introduced is called fault-tolerant. Such techniques will allow us to design networks built with noisy quantum gates that implement QECC fulfilling a requirement that may be roughly expressed as follows: a network is fault-tolerant (FT) [5] when it is able to eliminate more errors than those introduced by itself.

There are two important features that we need to keep in mind when we face the construction of fault-tolerant circuits [6]: firstly, quantum gates must be applied directly on the logical qubits, i.e. encoded (consequently avoiding the steps of decoding, gate application and encoding back again) and in the second place, the network must control the error spreading. All gates and measurements involved in the correction procedure of a QECC must fulfill both requirements. For a [[n,k,d]] stabilizer code, the error syndrome is obtained by means of a measurement of the (n-k) stabilizer generator operators, and these measurements must also be fault-tolerant. If not, error accumulation would rapidly destroy the coherence of the states in the computer. Shor and Steane propose [3] to carry out these measurements by copying the syndrome information of the state over some ancilla states. In both proposals the ancillas used are themselves coherent intricate states as is syndrome extraction.

In this paper we discuss up to which degree these complex ancillas are really advantageous compared to a simple ancilla with a non-coherent state followed by a non-fault-tolerant syndrome extraction. Simulations using several ancillas with a code [[7,1,3]] and similar codes are carried out in [7] and [8]. Instead of a direct code simulation as the present one, other authors have introduced the decoherence effect as an effective channel acting directly on the logical qubit. This method permits the threshold error probability to be calculated with fault-tolerant constructions in the the infinite concatenation limit [9].

## II. ERROR MODEL, ENCODING AND CORRECTION

We consider a specific symmetric physical qubit $|q\rangle = (|0\rangle+|1\rangle)/2^{1/2}$ exposed to a stochastic noise [10]. The model used introduces the error in the qubit by dividing the network into time steps and gates affecting only one or two qubits. The application of a gate may involve several time steps; in the model, the memory error of these steps is included in the time evolution between gates. Errors are introduced by means of an isotropic depolarizing channel model into which the evolution (or memory) error is made up by applying the operators $\sigma_X$, $\sigma_Y$ and $\sigma_Z$ (to abbreviate notation we will refer to them as X, Y and Z) with the same probability $\varepsilon/3$ (per time step and qubit), as long as the probability of having no evolution error is (1-$\varepsilon$). The error correction networks use gates affecting one qubit: Hadamard rotations (H), NOT gates (X) and measurements, each one having an error probability $\gamma$. The only gates applied affecting two qubits are CNOT gates, and their error is simulated through the set of 15 error possibilities of the set $\{I, \sigma_X, \sigma_Y, \sigma_Z\} \otimes \{I, \sigma_X, \sigma_Y, \sigma_Z\}$ each one having an error probability $\gamma/15$. In this way the error probability is larger than $O(\gamma^2)$, a fact that reveals an existing correlation between the errors in the qubits connected through the gate.

In order to control the induced decoherence, the qubit $|q\rangle$ is encoded by means of a stabilizer code, particularly a CSS (Calderbank-Shor-Steane) [11] code [[7,1,3]]. Every physical qubit is encoded in a logical qubit $|q_L\rangle$ involving seven physical qubits. The quantum encoding network appears in Figure 1.

This encoding network is not fault-tolerant, as it operates CNOT gates between the physical qubits making up $|q_L\rangle$. The criterion used to test the quality of the information recovery will be the fidelity of the output state $|q_{L,Corrected}\rangle$, generated once the error correction (and this constitutes the main objective) has been applied. Fidelity may be characterized by defining a fault path $F_p(\varepsilon,\gamma)$ [12] as the list

of locations, time steps and types of errors that take place during a complete calculation. Each fault path has a certain probability of occurrence $P(F_p)$ that depends on the number of errors involved, but not on the type of error, as all of them are equally probable and independent. A given path $F_p$ may be divided into two parts: $F_e$, corresponding to the encoding of the qubit, and $F_c$ depending on the correction network employed (and also on the ancilla), verifying $F_p(\varepsilon,\gamma) = F_c(\varepsilon,\gamma) \circ F_e(\varepsilon,\gamma)$. If $\rho$ is the density matrix for the initial state of the qubit, $A_i$ is the noise operator introducing an error at the i-th time step of the free evolution and $g_j$ is the operator introducing the error when a gate (o more) is applied at the j-th step, the effect of the fault path may be represented by an operator $Q(F_p)$ involving an encoding (E) and afterwards a recovery (R):

$$Q(F_p) = A_t \circ g_t \circ \cdots \circ A_1 \circ g_1 = R(F_c) \circ E(F_e)$$

with $A_i$, $g_j \in \bigotimes_{i=1}^{m} \{I, \sigma_x, \sigma_y, \sigma_z\}^{\otimes n}$, m being the number of qubits involved and n the record length. The final state of the qubit once it has crossed over the network may be expressed as the following average over all the fault paths:

$$Q(F_p) \circ \rho = \sum_{F_c, F_e} P(F_p) \{R(F_c) \circ E(F_e)\} \circ \rho$$

The numerical simulation carries out at least a number of calculations $N_C \sim 10 \max\{1/\varepsilon, 1/\gamma\}$, to determine finally the fidelity $F(\varepsilon,\gamma)$ as the following average:

$$F(\varepsilon,\gamma) = \langle q_L | Q(F_p) \circ \rho | q_L \rangle = \frac{1}{N_c} \sum_{i=1}^{N_c} |\langle q_{L,corrected,i} | q \rangle|^2$$

where the state $|q_{L,corrected,i}\rangle$ corresponds to the logical qubit after the correction applied in the i-th calculation and $|q_L\rangle$ is the initial encoded state without errors. Evidently, the quality of $|q_{L,corrected,i}\rangle$ depends on the values of $\varepsilon$ and $\gamma$. Errors are introduced into the calculation using the Luxury Pseudorandom Numbers [13] which is an improvement of the subtract-and-borrow random number generator proposed by Marsaglia and Zaman. The fortran-77 code is due to James [14], and is used with the luxury level parameter p = 223. As the code state for this value of p, any theoretically possible correlations have very small chance of being observed. The code returns a 32-bit random floating point number in the range (0, 1). For each run a new random seed is chosen as a 32-bit integer.

### A. Syndrome measurement and error correction.

The quantum code used has a distance 3, so it will be able to correct an error $\sigma_X$, $\sigma_Y$ or $\sigma_Z$ affecting any one of the seven physical qubits. An error that has possibly been introduced at the encoding step (not fault-tolerant) is determined by measuring the eigenvalues of the stabilizer code generators $S=\{S_i, i=1,...,6\}$

$$\begin{array}{ll} S_1 = (I\ I\ I\ Z\ Z\ Z\ Z) & S_4 = (I\ I\ I\ X\ X\ X\ X) \\ S_2 = (I\ Z\ Z\ I\ I\ Z\ Z) & S_5 = (I\ X\ X\ I\ I\ X\ X) \\ S_3 = (Z\ I\ Z\ I\ Z\ I\ Z) & S_6 = (X\ I\ X\ I\ X\ I\ X) \end{array}$$

thus obtaining the error syndrome that determines the structure of the correction operator. In a CSS code, the syndromes corresponding to bit-flip and phase-flip errors are separated, as the structure of the generators shows. Given a syndrome vector $(s_1, s_2, s_3; s_4, s_5, s_6)$ where $s_i \in \{0,1\}$, i=1...6, its first three components determine the bit-flip error (X), as long as the other three determine the phase-flip error (Z). The correction operator has the structure:

$$X_{(s_1,s_2,s_3)}Z_{(s_4,s_5,s_6)}$$

If the same position of the first and second cluster contains a 1, it indicates an error of type Y = XZ. The general correction scheme is shown in Figure 2 [15].

Several pieces in the network can be appreciated. The first one is the synthesis of the encoded qubit $|q_L\rangle$, then the ancilla-qubit interaction network (IN) allowing (through the suitable CNOT gates) the syndrome of the possible error to be copied from the qubit $|q_L\rangle$ to the ancilla states. The measurement of these states provides the six bits of syndrome. Finally, the qubit $|q_L\rangle$ is corrected and we calculate the fidelity $F(\varepsilon,\gamma)$ for $|q_{L,Corrected}\rangle$.

### B. Ancillas employed

We will follow different schemes to copy the error syndrome into the ancilla, depending on the degree of parallelism in the IN, the possible verification of the ancilla state and the syndrome repetition before acting on the qubit $|q_L\rangle$ to correct it. The different schemes discussed are detailed below.

#### 1. Simple ancilla.

The ancilla of IN-1 contains only three qubits; this is the minimum number necessary to extract the six classic syndrome bits, taking into account that the same three qubits are reused. This method offers the benefit of economy, and some of the difficulties are its small degree of parallelism and an ancilla not being fault-tolerant. Keeping in mind the parity check matrix of the classic code [7,4,3] it is easy to construct the network shown in Figure 3(a) (the meaning of the notation used in all the networks will be explained later, in the section *Recovery networks studied*).

Note the small degree of parallelism introduced into the application of the gates, as each time step (shown as a dotted vertical line) carries out three CNOT gates. The process of copying the syndrome into the ancilla takes 10 time steps. The last seven parallel Hadamard rotations (H circles) coincide in the same time step with the syndrome measurement gates (squares). The total network, in addition to 14 H gates, uses 24 CNOT gates.

During the operation time of the circuit, some errors may take place in the ancilla, resulting in an erroneous syndrome. A possible improvement would be obtained by measuring the syndrome an even number of times and choosing the most repeated one as the correct syndrome. In the first strategy the syndrome is obtained three times before deciding on the action for correcting $|q_L\rangle$. The network used is the same for a single syndrome measurement, but applied three times. Representing by only one CNOT gate with thicker lines every time step involving three parallel CNOT gates, the network used to measure the three syndromes may be represented as in Figure 3(b).

In this Figure 3(b), each ancilla state (represented by a horizontal dotted line) really corresponds to three qubits in an initial state $|000\rangle$. Every measurement (shown by a square) turns out three classic bits. The first part of the network provides three clusters characterizing the possible bit-flip error, while the second part reveals three more clusters that identify the possible phase-flip errors. The whole process involves 26 time steps, 72 CNOT gates, 14 H gates and 18 measurement gates.

Some modifications to this ancilla state have been discussed in order to improve the syndrome measurement process. The final purpose is to apply a fault-tolerant logic.

### *2. Shor's ancilla*

In order to bring the error spreading from ancilla to $|q_L\rangle$ under control and to increase the parallelism, this IN-2 (Figure 4) uses Shor's ancilla state, synthesized by rotating Hadamard (bit wise) a cat state (four qubits in an entangled state $(|0000\rangle+|1111\rangle)/2^{1/2}$). An error occurring in $|q_L\rangle$ ends up as a change of parity in Shor's ancilla state that, once detected, enables us to obtain a classic syndrome bit. The extraction of the six syndrome bits that characterizes bit-flip and phase-flip errors requires six time steps and 24 CNOT gates connecting the ancilla and $|q_L\rangle$. These CNOT gates are applied in parallel within six blocks including each one four gates.

The CNOT gates connecting $|q_L\rangle$ with the ancilla are carried out transversally. If an error happens in the ancilla, it will propagate only to one of the physical qubits that conform the encoded $|q_L\rangle$. A later correction would be able to detect and correct this error. The structure given to Shor's ancilla allows the spreading of errors affecting the ancilla to be controlled. Nevertheless, its synthesis circuit is not itself fault-tolerant since it involves CNOT gates between its physical qubits. For the ancilla, the probability of having two or more errors has an order of magnitude $O(\varepsilon,\gamma)$, and they will be propagated to $|q_L\rangle$ with the same probability. In order to control this possibility, Shor introduced a fifth qubit and two more CNOT gates as may be seen on Figure 4. When the measurement of this fifth qubit in the ancilla turns out to be a 1, the ancilla is rejected and a fresh one is synthesized. Now the probability for the cat state to contain bit-flip errors in two or more physical qubits will behave as $O(\varepsilon^2, \gamma^2)$, namely it is fault-tolerant for bit-flip errors.

Even though the total scheme is fault-tolerant for bit-flip errors, it is not so for phase-flip errors, since they are not detected at the fifth qubit introduced in the ancilla. Phase-flip errors will become bit-flip errors at Hadamard rotations applied at the end of the ancilla synthesis (when the syndrome of bit-flip errors in $|q_L\rangle$ is measured), generating a wrong syndrome. If we go on correcting with this incorrect syndrome, we will introduce new bit-flip errors in $|q_L\rangle$ that will be irrecoverable in later correction steps. Such a problem may be solved by repeating the syndrome extraction several (three) times, and choosing the most repeated one as the correct syndrome. Now the whole procedure is fault-tolerant. When three syndromes are measured, this network uses 72 CNOT gates connecting the ancilla and $|q_L\rangle$. As seen in Figure 4, H gates applied on $|q_L\rangle$ have been replaced by equivalent gates performed on ancilla state. This possibility would afford no benefit in the case of a simple ancilla IN-1.

### *3. Steane's ancilla.*

Steane's proposal (see Figure 5) starts with a $|0_L\rangle$ ancilla state synthesized by means of the network given in Figure 1 (but excluding the H gates in the last step). Two transversal CNOT gates (involving 14 CNOT between physical qubits) connect $|q_L\rangle$ with two ancillas in the state $|0_L\rangle$ and transfer the error syndrome to the ancillas. Their destructive measurement provides 14 classic bits in the form of two error vectors $\mathbf{e_x}$ and $\mathbf{e_z}$. The syndrome is obtained by means of $\mathbf{He_x} = (s_1, s_2, s_3)$ and $\mathbf{He_z} = (s_4, s_5, s_6)$. The process of copying the syndrome on the ancilla is implemented through two time steps by means of 14 CNOT gates. Its compact scheme is shown on Figure 5(a), whereas the ancilla synthesis network (including verification) appears in Figure 5(b).

In the schematic Figure 5(a), each horizontal line represents a logical qubit (encoded in seven physical qubits). Moreover, CNOT and Hadamard gates correspond to seven gates applied transversally, and being hence fault-tolerant.

Following the same ideas introduced for Shor's ancilla, an eighth qubit may be included (as shown in Figure 5(b) to make the detection of bit-flip errors in the ancilla state possible. When the logical value of this eighth qubit is 1, the ancilla is rejected and a fresh one synthesized. The eighth qubit cumulates the results of the checking bits corresponding to the classic code [7,3,4] with distance 4, enabling consequently errors with a weight w ≤ 3 to be detected. If the verification procedure operates without errors, the circuit detects all the bit-flip errors, as all errors with a weight w ≥4 are equivalent to errors with w ≤ 3 [16].

The complete fault-tolerance of the correction network is reached when the syndrome is calculated three times and the most repeated is taken as the correct one. In this case the interaction network involves 42 CNOT gates between $|q_L>$ and the ancilla (see figure 10 of [16]).

### *4. Parallelized Steane's ancilla.*

We have discussed the possibility of using an ancilla network with a higher degree of parallelization. Four new qubits are introduced to accumulate the bits of ancilla checking instead of only one (the eighth qubit of the preceding circuit). This construction allows us to parallelize the verification of the ancilla compared to the network IN-3, where 19 time steps were required. These four additional qubits bring the possibility of verifying the ancilla in only five time steps (using the same number of CNOT gates as in IN-3). This circuit, IN-4, is shown in Figure 6 (notice that the network appearing in [16] had some gates in an erroneous location).

### *5. Steane's ancilla with bit and phase-flip errors verification.*

Since the quality of the corrected qubit $|q_L>$ depends on how good the state of the ancilla is, we have finally proposed a network verifying not only the presence of bit-flip errors, but also phase-flip errors. The objective we are looking for is to provide an ancilla state as faithful as possible before letting it interact with the qubit to be corrected. The fraction of the network that verifies bit-flip errors is analogous to IN-4 (with four additional qubits as represented in Figure 6). The section of the network dedicated to verify phase-flip errors requires a Hadamard rotation (before and after the syndrome measurement) of the seven physical qubits encoding $|q_L>$ to transform it into the dual basis. Nevertheless, instead of applying 14 Hadamard gates to the ancilla, an operation that could introduce too many errors, the circuit is slightly modified (see Figure 7), applying H gates only to the three additional qubits (6 H gates in total), and the CNOT gates are inverted. Three syndrome bits are then obtained for the phase-flip errors, as the process works inside the code [7,3,4]. The three measurements taken just after the three H gates also serve to prepare the three $|0>$ states needed (besides a fourth one $|0>$) as input qubits for bit-flip error verification in the ancilla. Note that phase-flip error verification is carried out before bit-flip errors, because the last ones are more dangerous if they contaminate the qubit $|q_L>$ [16]. If any of the measurements (shown as M) detects a bit 1, the ancilla is rejected and a fresh one is synthesized. Figure 7 shows the ancilla synthesis network along with its verification step. This ancilla will be used in the IN-5.

### III. SIMULATION RESULTS

We have carried out a simulation for the correction process of errors introduced into a symmetric qubit $|q>=(|0>+|1>)/2^{1/2}$ encoded by means of a CSS code [[7,1,3]] and using the different ancillas detailed above. In all cases the correction considers the possibilities of measuring only one or three syndromes (choosing the most repeated one), and also the possibility of verifying or not the state of the ancilla before carrying out the error correction. All these eventualities are discussed as a function of the evolution error ε and gate error γ. The results of the simulation appear in Figures 8, 9 and 10.

#### **A. Recovery networks studied.**

To facilitate the display and comparison of the results obtained, we summarize now the notation followed for the different ancillas and correction procedures applied:

IN-1 (1,0) = Simple ancilla + 1 Syndrome + Without ancilla verification (0).
IN-1 (3,0) = Simple ancilla + 3 Syndromes + Without ancilla verification (0).

IN-2(i,0) = Shor's Ancilla + i Syndromes + Without ancilla verification (0).
IN-2(i,V) = Shor's Ancilla + i Syndromes + With ancilla verification (V).

IN-3(i,0) = Steane's Ancilla + i Syndromes + Without ancilla verification (0).
IN-3(i,V) = Steane's Ancilla + i Syndromes + With ancilla verification (V).

IN-4(i,0) = Steane's Ancilla parallelized + i Syndromes + Without ancilla verification (0).
IN-4(i,V) = Steane's Ancilla parallelized + i Syndromes + With ancilla verification (V).

IN-5(i,V) = Steane's Ancilla parallelized with bit and phase-flip errors verification + i Syndromes + With ancilla verification (V).

Whenever we refer to the ancilla-qubit interaction network, the notation IN-n(i,x) is used, with n=1,2,..5, x = 0,V and i = 1, 3, whereas an(i,x) appears when we are only interested in mentioning the type of ancilla used.

Determined to make the comparison of the results obtained easier (Table I), the different IN are classified in terms of the following variables:

T = Number of time steps needed to synthesize the ancilla + those belonging to the interaction ancilla-$|q_L>$. The final time step representing the qubit $|q_L>$ correction is not counted up.

G = Total number of gates applied in the ancilla-$|q_L>$ interaction network, including those gates affecting only one qubit (H and measurements) as well as those involving two qubits (CNOT). The NOT gates necessary to correct $|q_L>$ are not counted.

Q = Physical qubits required in the ancilla, together with those needed by its verification.

TABLE I. Description and characteristics of ancilla's networks.

| IN-1 (Simple ancilla) | 1(1,0) | 1(3,0) |
|---|---|---|
| T | 10 | 26 |
| G | 44 | 104 |
| Q | 3 | 18 |

| IN-2 (Shor) | 2(1,0) | 2(1,V) | 2(3,0) | 2(3,V) |
|---|---|---|---|---|
| T | 11 | 14 | 23 | 26 |
| G | 96 | 114 | 288 | 342 |
| Q | 24 | 30 | 72 | 90 |

| IN-3 (Steane) | 3(1,0) | 3(1,V) | 3(3,0) | 3(3,V) |
|---|---|---|---|---|
| T | 12 | 31 | 17 | 36 |
| G | 66 | 104 | 198 | 312 |
| Q | 14 | 16 | 42 | 48 |

| IN-4 (Steane's parallelized) | 4(1,0) | 4(1,V) | 4(3,0) | 4(3,V) |
|---|---|---|---|---|
| T | 7 | 12 | 11 | 16 |
| G | 66 | 104 | 198 | 312 |
| Q | 14 | 22 | 42 | 66 |

| IN-5 | 5(1,V) | 5(3,V) |
|---|---|---|
| T | 18 | 22 |
| G | 146 | 438 |
| Q | 22 | 66 |

The starting point for the discussion is to contrast the behavior of the infidelity (1- $F(\varepsilon,\gamma)$) as a function of the evolution error $\varepsilon$ (per qubit and time step) and the gate error $\gamma/7$ for the cases of a simple ancilla 1(1,0) and Shor's ancillas 2(1,0) and 2(1,V). The infidelity is chosen instead of the fidelity itself, because the first shows more clearly the quality behavior in the small $\varepsilon$ or $\gamma/7$ values, especially when a logarithmic scale is used. Beginning with these two networks we will discuss the modifications that improve the structure of the ancilla attending to its parallelism, transversality in the ancilla-$|q_L\rangle$ interaction and measurement of several syndromes before the correction. This last option belongs to a full fault-tolerant method. The cases studied include the $\varepsilon$ dependence of the infidelity for constant $\gamma = 0.001$ and the $\gamma$ variation maintaining fixed $\varepsilon = 0.001$. In all the cases studied the results show a more market dependence of $\gamma$. The infidelity spreads quickly in the case of $\varepsilon$ constant (0.001) and $\gamma$ decreasing.

### B. Effect of the different ancillas.

Having compared all the results obtained with one syndrome, in general Shor's ancilla's 2(i,x) provides the highest infidelity values and their relative behavior, with regard to the other ancillas, do not seem to change when $\varepsilon$ or $\gamma$ varies. Their IN-2 uses the greatest number of gates connecting $|q_L\rangle$ with the ancilla (24 CNOT gates), in spite of the property that ancilla states are not very complicated (see figure 4). Remark that, although the total number of gates in circuits 5(1,V) and 5(3,V) is higher, only 14 of them connect $|q_L\rangle$ with the ancilla. This fact explains why in this case the infidelity is much smaller than the one obtained with Shor's ancillas.

If we just pay attention to networks involving only one syndrome measurement and without ancilla verification, those IN-n(1,0) having n = 1,2,3 and 4 (maybe excepting ancilla 5) involve a similar number of time steps (T), and the best IN is the one that uses the lowest number of gates (G) in contrast to the conclusion reached in [8]. The best ancilla is the simplest one 1(1,0), using the smallest number of qubits (three) and gates, as it requires no gate to prepare the initial $|000\rangle$ state. When $\varepsilon$ varies (figure 8a, $\gamma=0.001$), the worst ancilla is Shor's (2(1,0)), with a high number of gates and qubits. If we consider 3(1,0) and 4(1,0), the second one is the best, because it needs a smaller number of time steps in the ancilla synthesis. Notice the network 5(1,0) would be equal to 4(1,0), so it is not considered. The cross of the curves 1(1,0) and 4(1,0) (figure 8a) for $\varepsilon \sim 3 \cdot 10^{-3}$ is very interesting. Surely it is a consequence of the fact that in the second ancilla the interaction ancilla-$|q_L\rangle$ takes place in a transversal way (more tolerant to error spreading) introducing hence a minor number of errors in $|q_L\rangle$.

When $\varepsilon$ is constant the infidelity variation with $\gamma$ is more evident, but the previous relative behaviour is maintained. A strong dependence of the infidelity with $\gamma/7$ is observed for the 4(1,0) network because it has a medium number of gates G (66) in conjunction with the smallest time steps T (7). This last fact causes the crossing with the 1(1,0) infidelity curve at $\gamma/7 \sim 10^{-4}$, improving the results for smaller error rate.

So looking at figures 8a and 8b, the best ancilla (if only one syndrome is measured when correcting $|q_L\rangle$ and no verification is included) is the simplest 1(1,0) (involving smaller G and Q), except when the gate error is small enough so that the ancilla parallelization turns out to be important. In this case the 4(1,0) ancilla network provides better results.

### C. Effect of ancilla verification

If we restrict the measurement of only one syndrome, the inclusion of ancilla's verification is not beneficial until one gets to values of $\varepsilon$ small enough. When $\gamma = 0.001$ (figure 8a, represented as a function of $\varepsilon$) ancilla 1(1,0) provides better fidelity values than ancilla 2(1,0) for all the $\varepsilon$ range considered. Such behavior is not surprising, as 2(1,0) needs a greater number of gates and qubits (see the previous section and table I). An ancilla including a verification step, as is 2(1,V) would improve the syndrome quality, including a fifth qubit (as shown in figure 4). Two CNOT gates deal with the state verification; their control bits are the first and fourth (respectively) and their image bit is the fifth one. If a bit-flip error occurs during the ancilla synthesis, the measurement of the fifth qubit will have the logical "1" value; consequently the ancilla will be rejected and a fresh one synthesized. Figure 8a shows the effect of such ancilla verification on the infidelity, plotted as a function of $\varepsilon$. We see that IN-2(1,0) is better than IN-2(1,V), except for $\varepsilon < 10^{-4}$. Values of $\varepsilon > 10^{-4}$, are not small enough as to compensate the error included in the verification step, related to an increasing in the number of gates and time steps. The 3(1,V) and 4(1,V) have a similar behavior, showing some benefits for $\varepsilon < 10^{-4}$. The inclusion of a phase error verification step in ancilla (5(1,V)) before it interacts with $|q_L\rangle$ seems to be advantageous respect to the 4(1,V) ancilla, when $\varepsilon < 4 \cdot 10^{-3}$, condition under which it improves the results obtained with ancilla 4, although the infidelity for 5(1,V) continues being greater than the values for 1(1,0).

Infidelities shown in figure 8b as a function of $\gamma/7$ ($\varepsilon = 0.001$) keep their relative positions when the verification is included or not. All of them show a better behavior when no ancilla verification is included. In contrast to their variation with $\varepsilon$, when $\gamma/7$ decreases, the fidelities obtained with ancillas an(1,0) and an(1,V) (an = 1,2,3,4) not only do not cross but moreover they separate each other. The greater number of gates, time steps and qubits used does not compensate for the benefit derived from the verification of the ancilla quality before letting in interact with $|q_L\rangle$ to obtain the syndrome. Hence, if we include ancilla verification we must be very careful in designing the network if we want to achieve benefits. Notice that the infidelity for ancilla 5(1,V) is completely analogous to the ancilla 4(1,V). Phase error verification provides no advantage for $\varepsilon = 0.001$ and the $\gamma/7$ values considered. The crossing between the 2(1,0), 2(1,V) and 3(1,V) at $\gamma/7 \sim 3 \cdot 10^{-4}$ reflects the importance of the evolution error through the T values shown in table I. The first one 3(1,V) has the biggest T, so it shows a higher infidelity.

When using only one syndrome, there is no difference between ancillas 4(1,V) and 5(1,V) (note that, if there is no verification, both ancillas are the same). No benefit is observed due to the inclusion of phase error detection in the ancilla, at least in the range of $\varepsilon$ values considered (maybe there will be some differences for $\varepsilon$ and $\gamma/7$ out of the range considered). The advantage reached for phase error checking is lost because the number of gates involved increases (see table I, IN-4 and 5).

Anywhere, for $\gamma = 0.001$, and as $\varepsilon$ decreases ( $< 0.001$) we can see an improvement in the fidelity of the ancilla 5(1,V) compared to 4(1,V). This feature is a consequence of the fact that $\gamma$ is sufficiently small and the reduction of the evolution error succeeds in compensating the higher number of gates in ancilla 5.

It is surprising, once again, that the simple ancilla 1(1,0) produces the best fidelity among those simulated, except in the case 4(1,0) for $\gamma/7 < 10^{-4}$ (where $\varepsilon = 0.001$) and for $\gamma = 0.001$ and $\varepsilon > 0.003$. It is precisely in this area of $\gamma/7$ values where the transversality of the interaction between ancilla 4 and $|q_L>$, as well as the parallelism in the ancilla verification, begins to generate benefits.

A previous analysis leads to the conclusion that, when only one syndrome is measured, the best ancillas are the simplest ones, and the inclusion of a verification step of bit-flips or phase-flips does not represent a clear advantage, except (perhaps) for very small error rate not shown in the figures.

### D. Effect of the parallelism in the ancilla

The effect produced by the parallelism in the ancilla's synthesis circuit may be discussed by comparing ancillas 3 and 4, since they use the same number of gates. The network referred as IN-4(1,V) uses less time steps (with a parallelism ratio for the synthesis and verification of the ancilla $R_p(4(1,V))$ = Gates/Time steps in ancilla = 31/9 = 3.44) than network 3(1,V) (with a parallelism ratio $R_p(3(1,V))$ = 31/28 = 1.11). The effect of the reduction in the number of time steps makes evident when we look at the dependence of fidelity as a function of $\varepsilon$ (figure 8a) or of $\gamma/7$ (figure 8b).

### E. Effect of syndrome repetition

The syndrome repetition maybe the most decisive fact to be considered before deciding what kind of ancilla must be used. Until now we have repeated the syndrome three times, and have chosen the most repeated one before correcting $|q_L>$. Whereas including an ancilla verification step or not, or introducing a higher degree of parallelism in ancilla's synthesis network doesn't seem to be, by themselves, the essential elements responsible for an improvement in the fidelity when it is compared to a simple ancilla 1(1,0) with only one syndrome calculated, the repetition of the syndrome is able to achieve this improvement (figures 9a and 9b).

The inclusion of a step devoted to the ancilla's verification provides a clear benefit only when the syndrome is repeated. In this case, the whole correction procedure is fault-tolerant. If the syndrome is repeated three times, those ancillas that include verification provide the best fidelities. Anyway, the fact that the method, including ancilla verification and syndrome repetition is fault-tolerant will not guarantee that the results obtained will be better than those obtained with a simple ancilla. In fact, among all the ancillas considered in this work, only numbers 4 and 5, i.e. the most parallelized and verified, provide similar fidelities than 1(1,0). Obtaining three syndromes by means of a simple ancilla 1(3,0) worsens the fidelity due to the large number of time steps used in the $|q_L>$-ancilla interaction. Infidelities obtained with ancillas 4 and 5 are totally similar (at least for the range of $\varepsilon$ and $\gamma$ values studied) and including a phase-flip error verification step in the ancilla doesn't seem to provide a remarkable advantage.

It is strikingly surprising that a simple ancilla with only one syndrome and without verification (1(1,0)) provides better fidelities than the fault-tolerant IN-n(3,V) with n = 2 and 3, and similar results for those obtained with n = 4 and 5. This difference is mainly due to the better parallelism ratio ($R_P (4(i,V))$ = 31/9 = 3.44 whereas $R_P (3(i,V))$ = 31/28 = 1.1) even though it needs to introduce four additional qubits to verify the ancilla.

So the question now would be could we do something to improve the results making the power of a fault-tolerant error correction method evident? One final improvement in the strategy would be not always to repeat the syndrome three times. Instead we calculate it two times and if they agree, we use it to correct the $|q_L>$ state. Otherwise, a third syndrome is calculated and the most repeated one is taken to correct the logical qubit. In the case of three different syndromes, no action is carried out in order not to make the qubit quality worse. So that the total number of time steps and gates per error correction is in average less or equal to the used in the three syndrome strategy, so we hope to improve the results. New infidelities obtained are plotted in figure 10. For $\gamma = 0.001$ and $\varepsilon$ varying, the infidelities for an(3,V) (an = 2, 3, 4 and 5) are smaller than those obtained using always three syndromes, and converge for small enough $\varepsilon$ values. The 2(3,V) provides a slightly higher infidelity than the rest (an(3,V), an = 3,4,5), reflecting a larger number of CNOT gates connecting the ancilla with the qubit, 72 for the first whereas 42 for the later.

## IV. CONCLUSIONS

We have studied the process of error correction applied to a symmetric qubit exposed to decoherence that is simulated through an error model of depolarizing isotropic channel. The quality criterion considered for the correction is the fidelity as a function of evolution errors in the range $10^{-4} \leq \varepsilon \leq 0.01$ and gate errors $3.10^{-5} \leq \gamma/7 \leq 0.01$. The networks discussed use five different ancilla states: simple ancilla, Shor's, Steane's, Steane's parallelized and Steane's parallelized with bit and phase-flip errors verification. Furthermore, they also include, when possible, the measurement of one or three syndromes, ancilla's verification and different degrees of parallelism. The number of different correction networks studied is 16.

The results obtained show that, in the case where only one error correction is applied and only one syndrome is measured, the best fidelity is provided by a simple ancilla without verification, at least in the range of errors studied. Even if the method used is fault-tolerant, the crucial element of the correction is the syndrome repetition, allowing the most repeated one to be chosen before carrying out the correction. In this way, errors affecting two or more physical qubits in $|q_L>$ (therefore irrecoverable) have a probability $O(\varepsilon^2, \gamma^2)$. Anyway, syndrome repetition is not enough to obtain the best results. It must be applied together with an ancilla verification involving, itself, a highly parallelized circuit. All these requirements compel us to introduce a large amount of physical qubits, and for this reason the error correction procedure is very expensive. Surprisingly, we see that a method using only three physical qubits provides not too bad results bearing in mind its economy. This result will have to be borne in mind when manufacturing quantum computers with a size not too large due to additional circuits dedicated to decoherence control.

The use of a fault-tolerant error correcting method repeating the syndrome three times seems to be quite expensive when only one correction is made. To improve the results in order to show the advantage of a fault-tolerant error correction, we have studied a modification in the strategy. The starting point calculates two syndromes, and if they agree it is used to correct the qubit, otherwise a third one is obtained and the most repeated one is used in the correction. In some cases the three syndromes are different, and we do not take any action with the intention to not decrease the qubit quality. With this strategy the fidelities show an improvement respect to those obtained with the three syndrome repetition method.

Finally we note that the conclusions reached are restricted only to those ancillas studied, error intervals indicated, the model developed and, maybe this is the most important restriction, to perform only one correction in time. Perhaps time evolution will imply certain differences concerning the adequacy of different ancillas to control an excessive error accumulation. We expect that this dangerous behavior will appear in the case of a simple ancilla. Fidelity may become zero if one or more (two) errors appear in $|q_L>$. In the first case, a subsequent correction may eliminate the error but not in the second case. This kind of behavior cannot be appreciated with only one correction. Nevertheless, we hope that the conclusions reached on the relative quality of the remaining ancillas will hold when the correction procedure will be applied through several time steps.

## ACKNOWLEDGEMENTS.

This work was supported by Plan Nacional I+D+I, Project BFM 2002-01414.

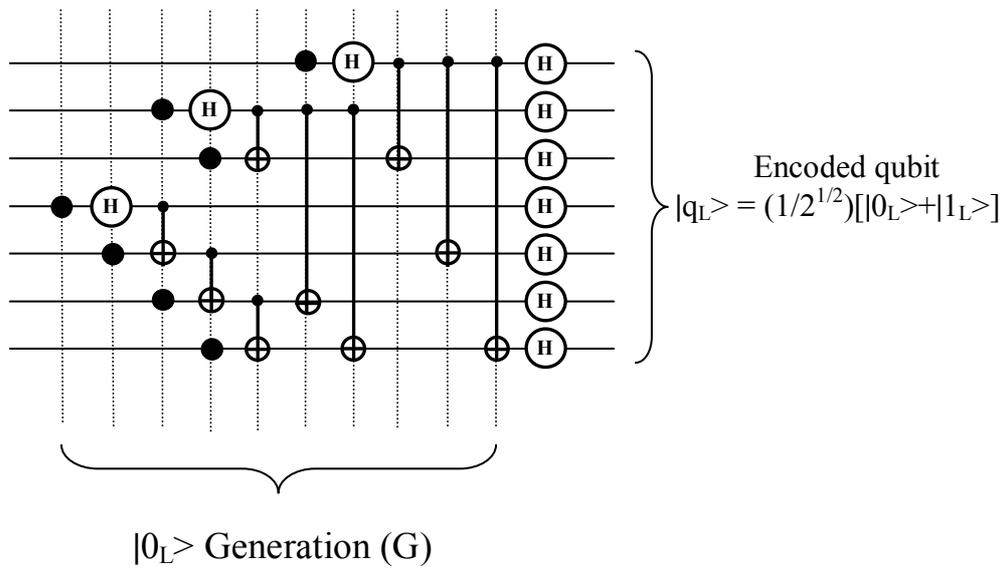

FIGURE 1. Qubit encoding network producing $|q_L\rangle$. Black dots represent physical $|0\rangle$ qubits, H are Hadamard gates and CNOT gates have their usual symbol. Vertical dotted lines mean time steps and horizontal lines are qubit evolution in time.

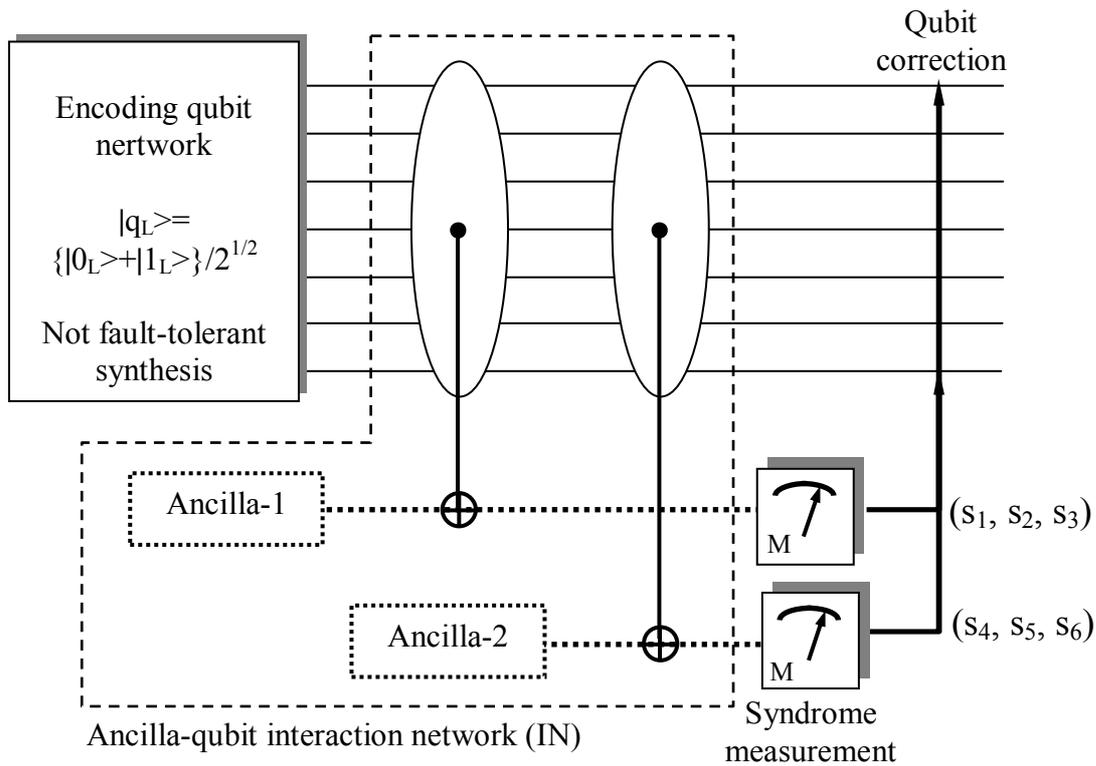

FIGURE 2. General error correction scheme.

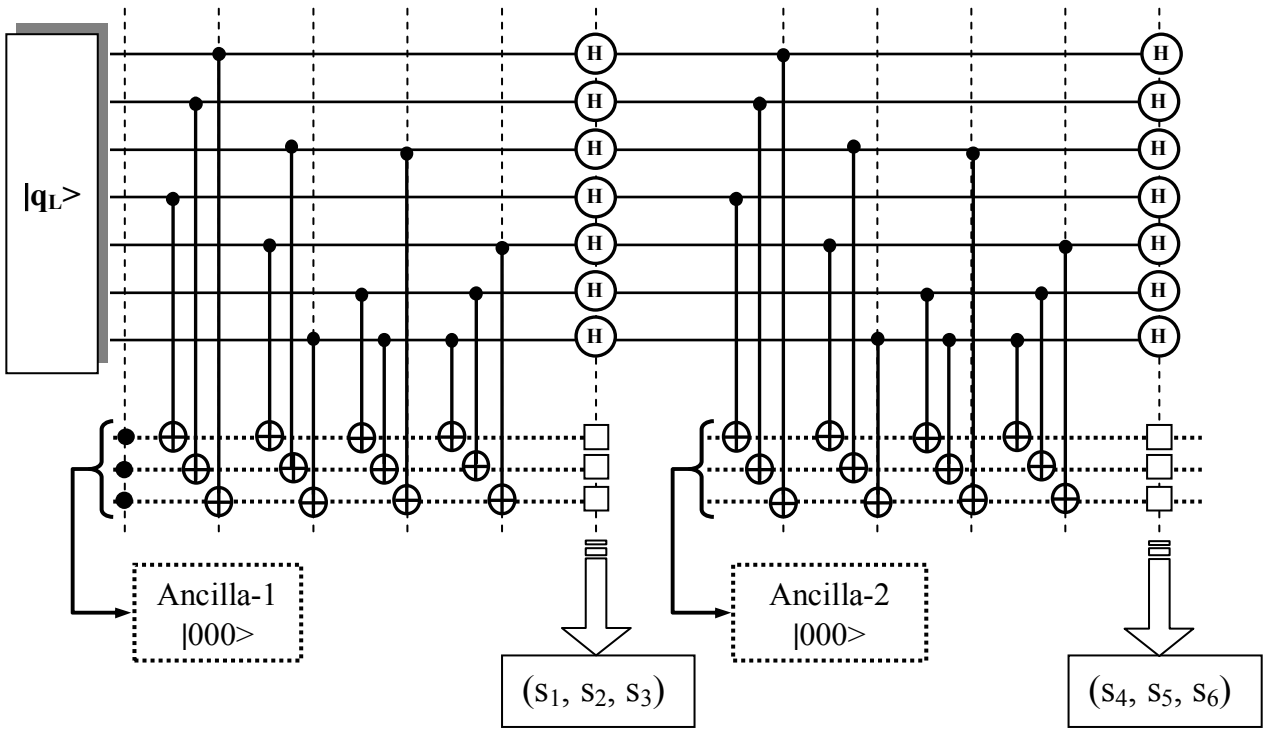

FIGURE 3a. Simple ancilla-qubit interaction network, measuring one syndrome (IN-1 (1,0)).

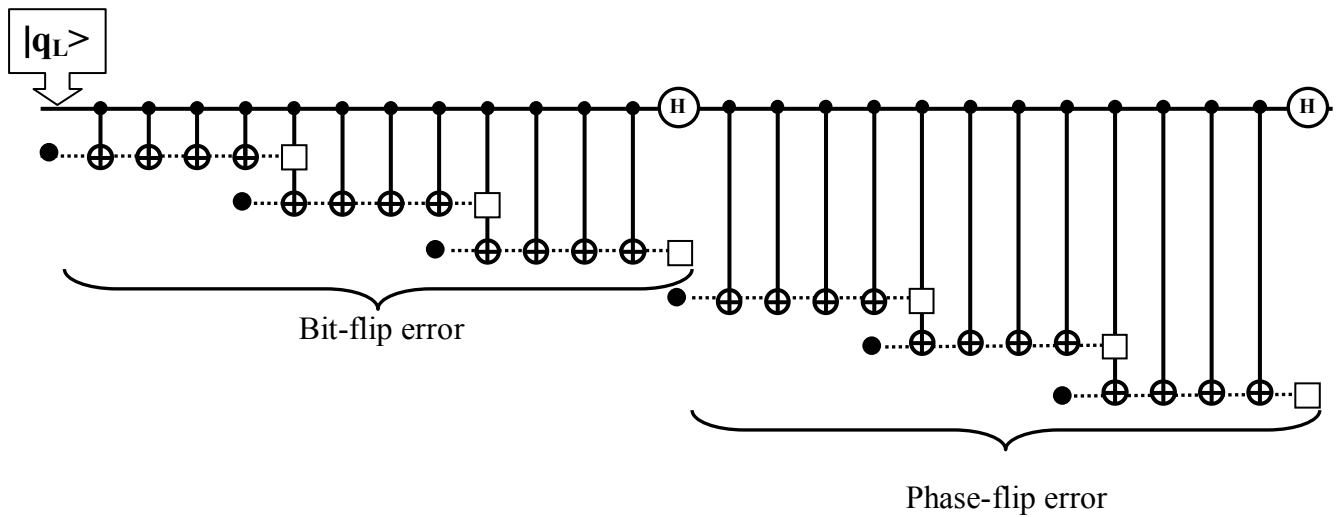

FIGURE 3b. Simple ancilla-qubit interaction network, measuring three syndromes (IN-1 (3,0)). Each thick CNOT gate represent three two-qubit CNOT gates applied in parallel.

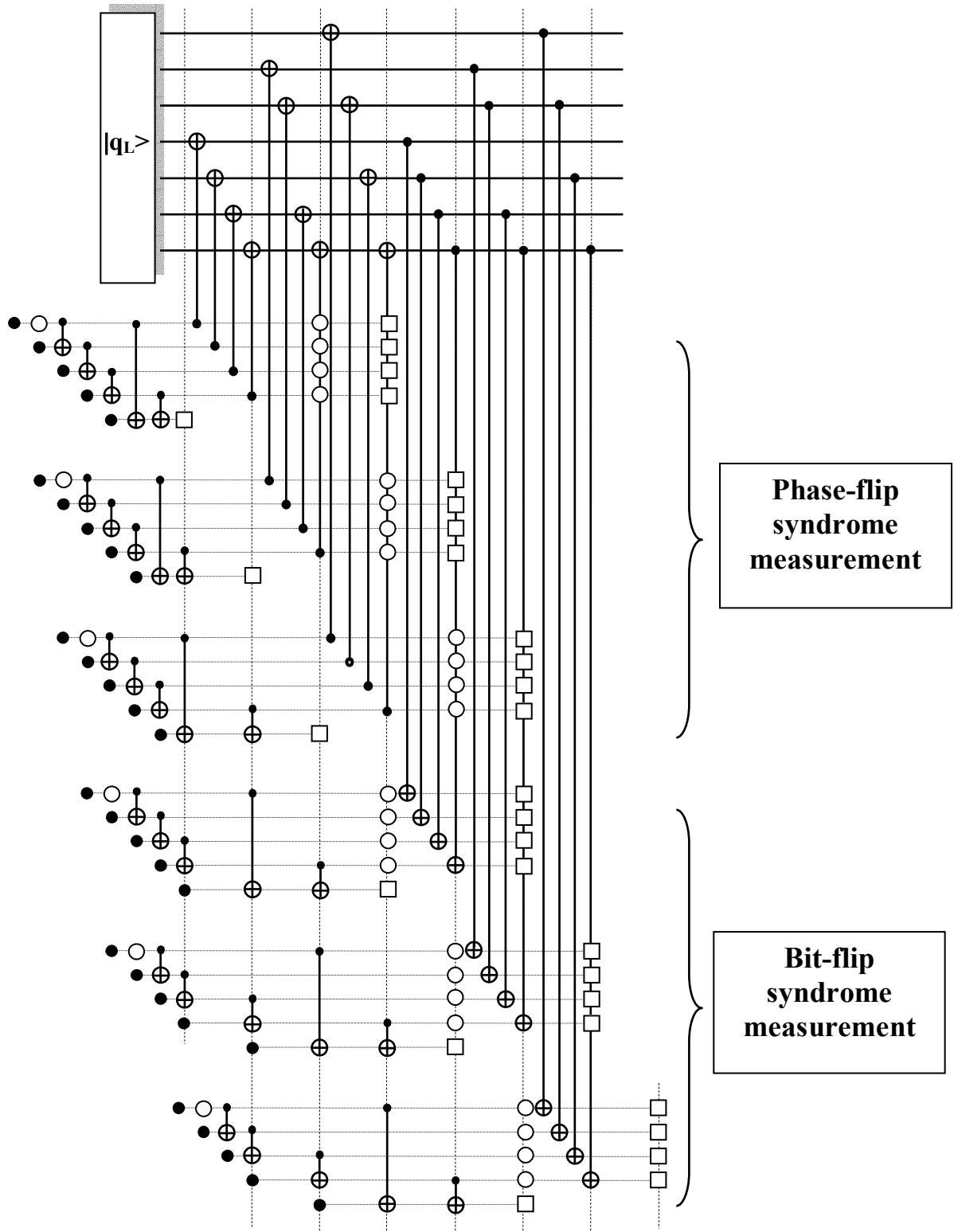

FIGURE 4. Detailed part of the ancilla-qubit interaction network for syndrome extraction using Shor's ancilla (IN-2(1,V)). When no ancilla verification step is included, the fifth ancilla qubit disappears.

**(a)**

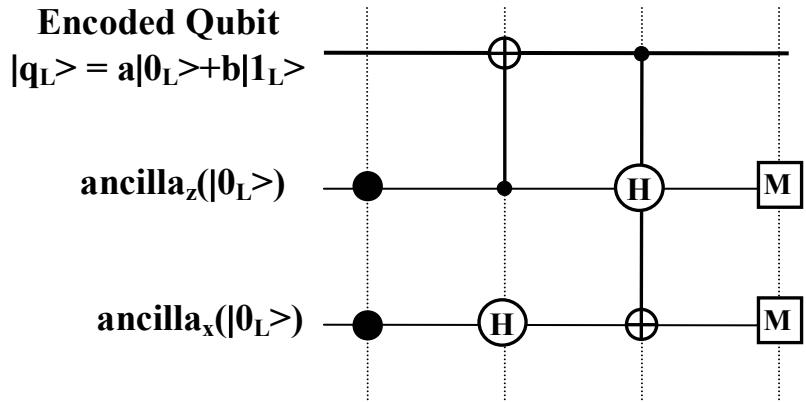

FIGURE 5a. Schematic correction network for IN-3, 4 and 5 (in compact notation). The CNOT, H and measurement gates represent seven transversal gates and black dots are encoded $|0\rangle$ states.

**(b)**

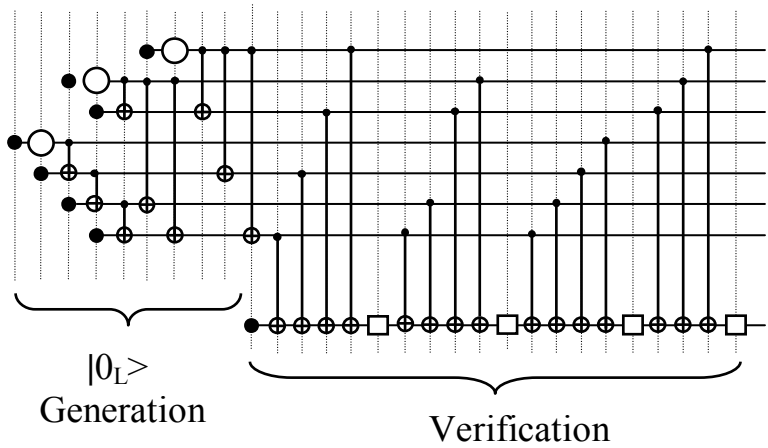

FIGURE 5b. Steane's ancilla network including the verification step used in the IN-3.

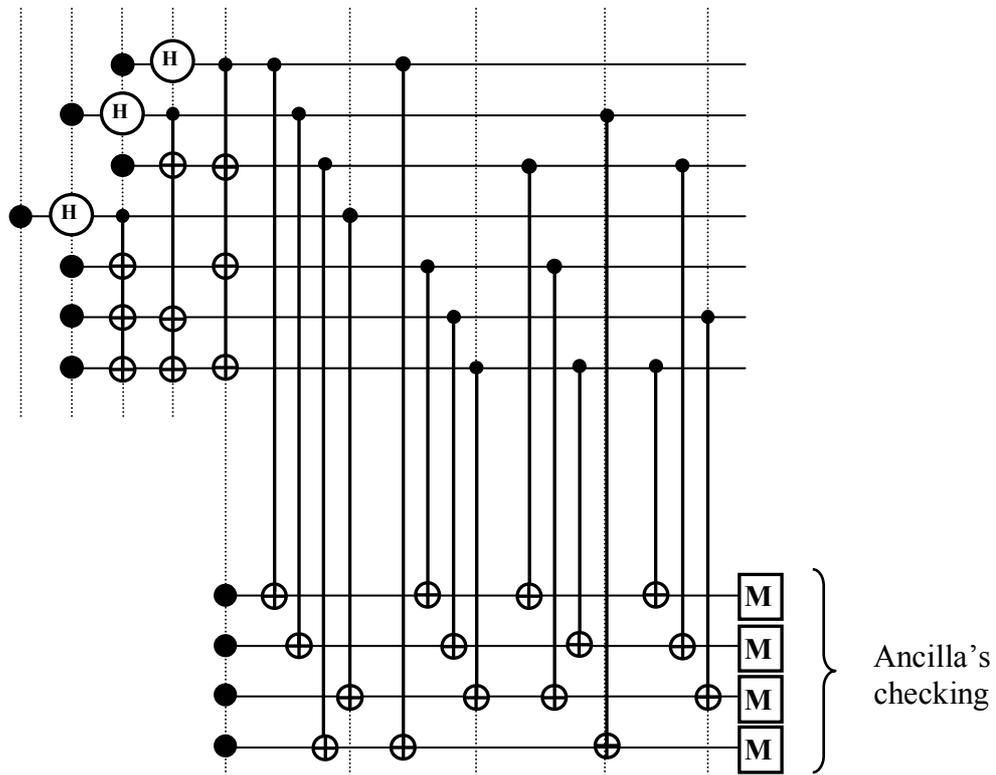

FIGURE 6. Parallelized Steane's ancilla network used in the IN-4.

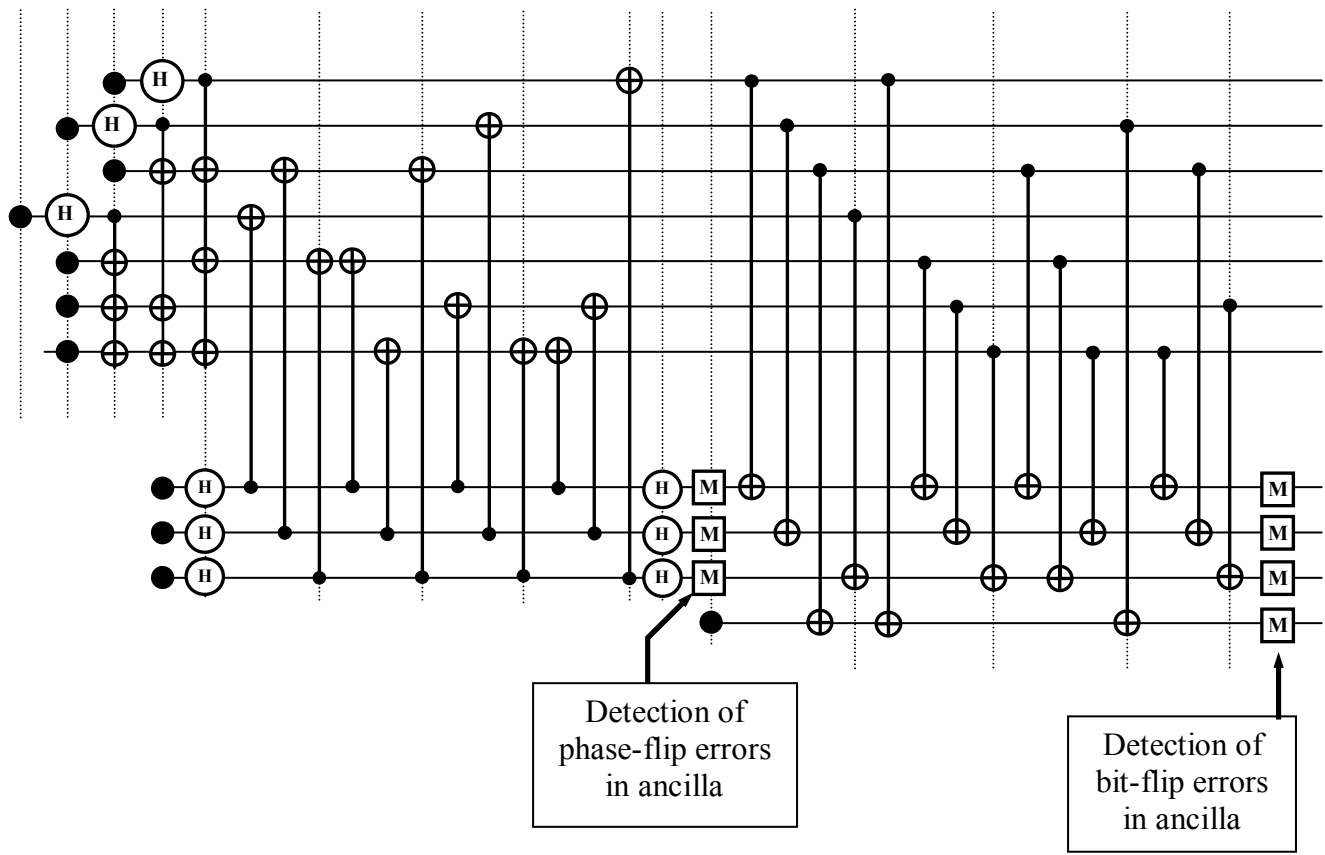

FIGURE 7. Steane's ancilla network including bit and phase-flip error verification used in the IN-5.

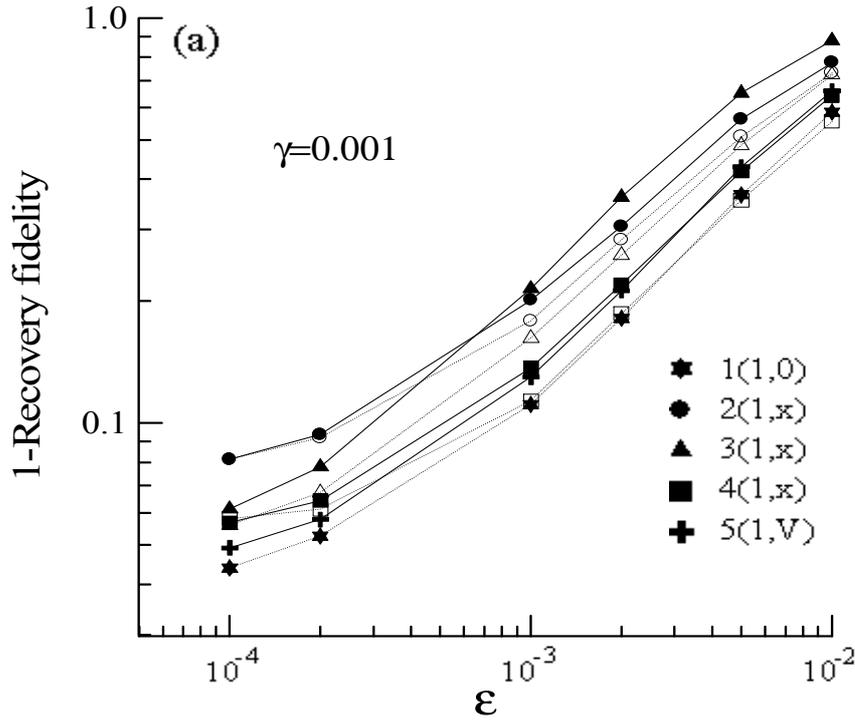

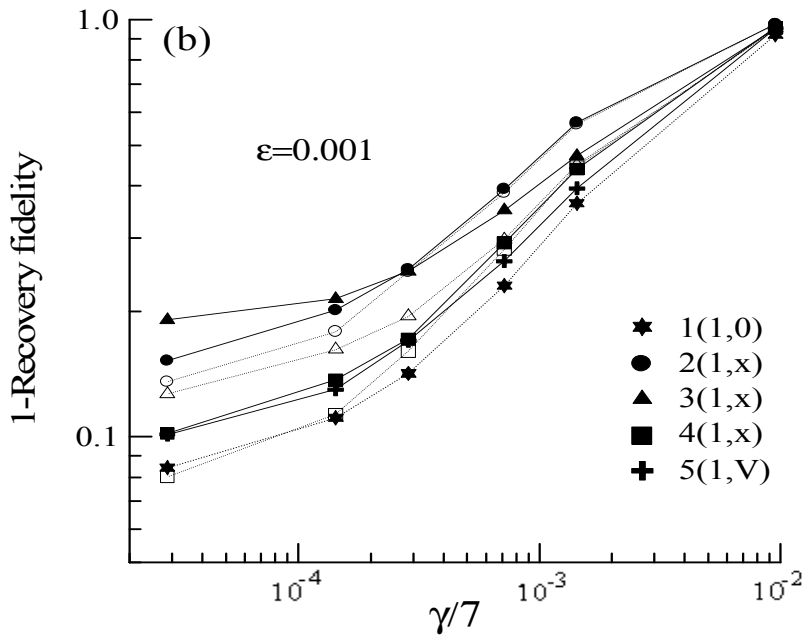

FIGURE 8. Infidelity (1-F($\varepsilon,\gamma$)) for the whole process after the qubit $|q_L\rangle$ correction for one syndrome measurement, (a) as a function of the evolution error $\varepsilon$ (with $\gamma = 0.001$) and (b) as a function of the gate error $\gamma/7$ (with $\varepsilon = 0.001$). Continuous lines (with full symbols) include ancilla verification (x = V) and dashed lines (open symbols) do not (x = 0).

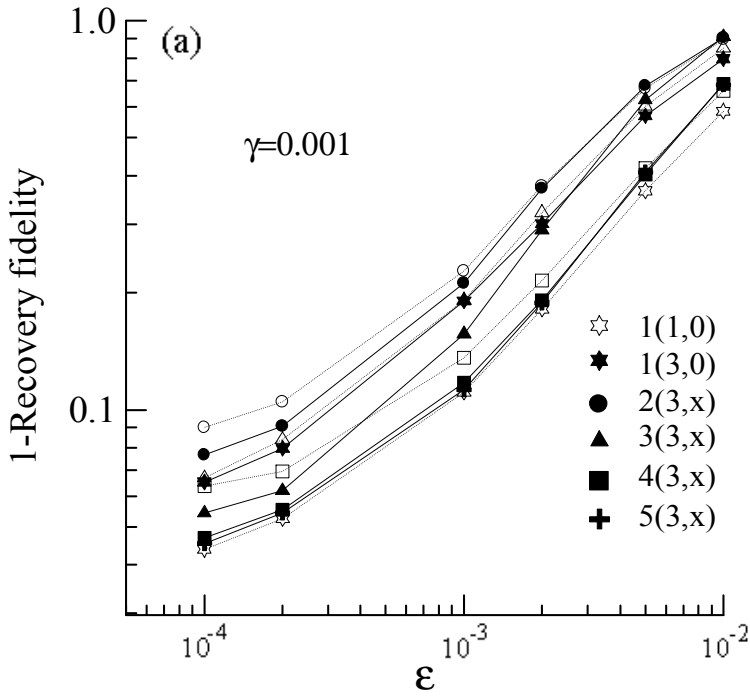

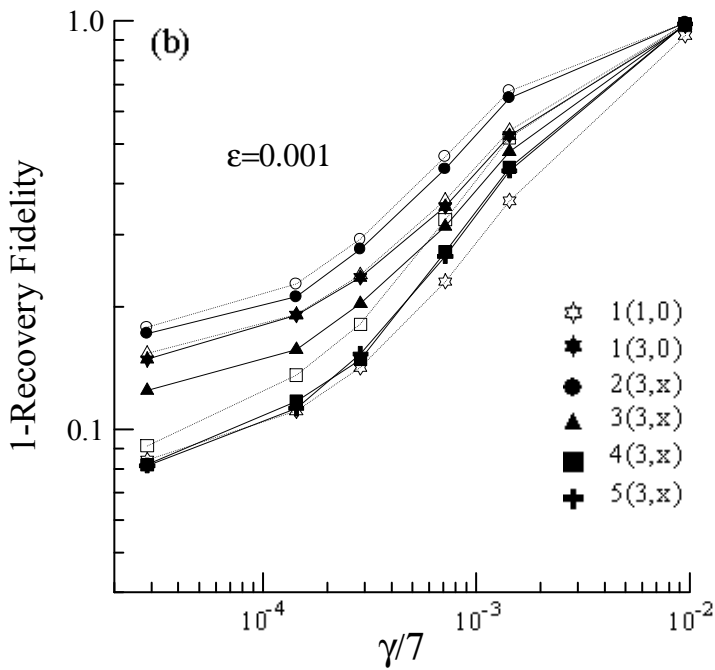

FIGURE 9. Infidelity $(1-F(\varepsilon,\gamma))$ for the whole process after qubit $|q_L\rangle$ correction with three syndromes measured, (a) as a function of evolution error $\varepsilon$ (with $\gamma = 0.001$) and (b) as a function of the gate error $\gamma/7$ (with $\varepsilon = 0.001$). Continuous lines (with full symbols) include ancilla verification (x = V) and dashed lines (open symbols) do not (x=0).

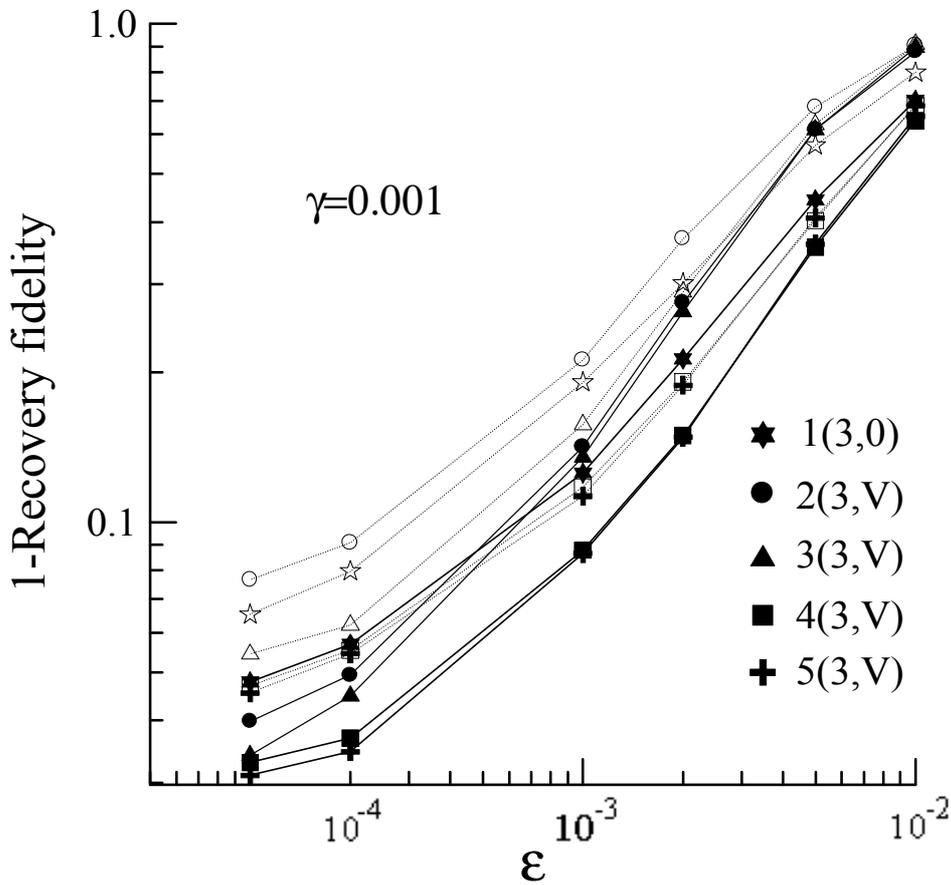

FIGURE 10. Infidelity (1-F($\varepsilon,\gamma$)) vs. $\varepsilon$ for $\gamma = 0.001$, for the whole process after qubit $|q_L\rangle$ correction by means of two different strategies: dashed lines (with open symbols) the method always calculate three syndromes before the qubit correction and continuous lines (with full symbols) first two syndromes are obtained and if they do not agree, a third one is calculated.